# ZERO POWER CHEMICAL SENSORS

Aishwaryadev Banerjee

We present a new class of batch-fabricated, low-power and highly sensitive chemiresistive sensors. We first present the design, fabrication, and characterization of batch-fabricated sidewall etched vertical nanogap tunneling-junctions for bio-sensing. The device consists of two vertically stacked gold electrodes separated by a partially etched sacrificial spacer-layer of α-Si and $SiO_2$. A ~10 nm wide air-gap is formed along the sidewall by a controlled dry etch of the spacer, whose thickness is varied from ~4.0 – 9.0 nm. Using these devices, we demonstrate the electrical detection of certain organic molecules from measurements of tunneling characteristics of target-mediated molecular junctions formed across nanogaps. When the exposed gold surface in the nanogap device is functionalized with a self-assembled monolayer (SAM) of thiol linker-molecules and then exposed to a target, the SAM layer electrostatically captures the target gas molecules, thereby forming an electrically conductive molecular bridge across the nanogap and reducing junction resistance. We then present the design, fabrication and response of a humidity sensor based on electrical tunneling through temperature-stabilized nanometer gaps. The sensor consists of two stacked metal electrodes separated by ~2.5 nm of a vertical air gap. Upper and lower electrodes rest on separate 1.5 µm thick polyimide patches. When exposed to a humidity change, the patch under the bottom electrode swells but the patch under the top electrode does not, and the air gap thus decreases leading to increase in the tunneling current across the junction. Finally, we present an electrostatic MEMS switch which is triggered by a very low input voltage in the range of ~50mV. This consists of an electrically conductive torsional see-saw paddle with four balanced electrodes. It is symmetrically biased by applying the same voltage at its inner electrodes leading to bistable behavior at flat or collapsed equilibrium positions. The use of elevated symmetric bias softens the springs such that the paddle collapses when a few milliVolts are applied to one of its outer electrodes thus causing the device to snap in and result in switch closure. Using the "spring softening" principle, we also present an application of a new kind of high sensitivity chemo-mechanical sensors.

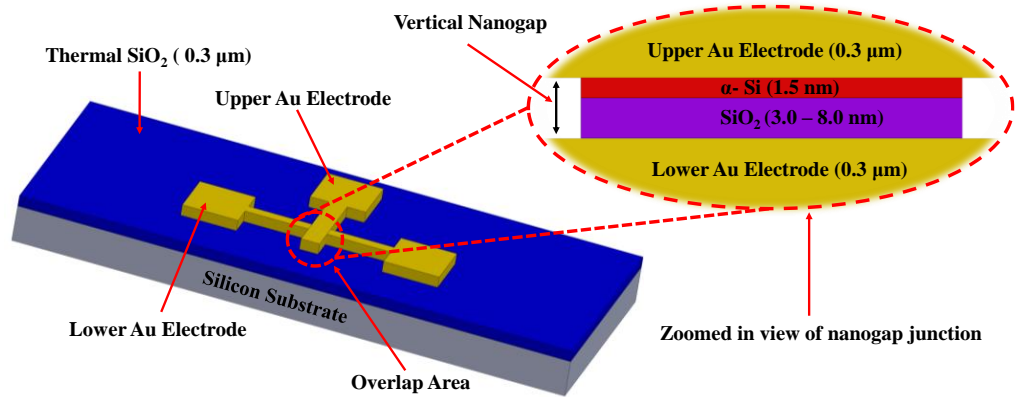

**Figure 2.1**: 3D schematic of device of vertical nanogap structure and zoomed in view of sacrificial spacer layer.

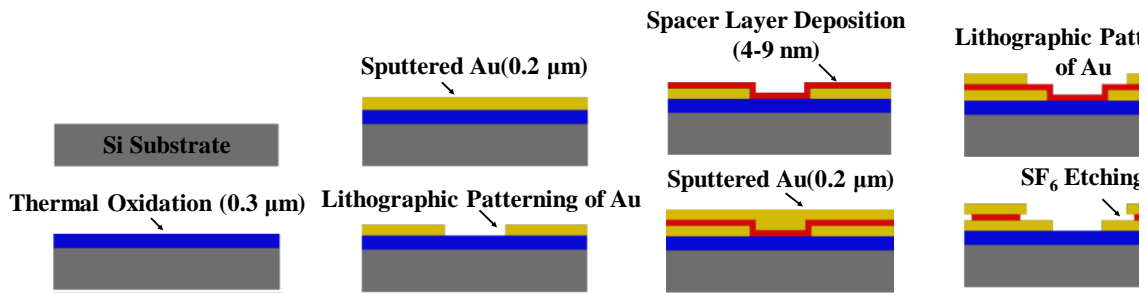

**Figure 2.2**: Simplified fabrication process of vertical nanogap electrodes separated by a thin spacer layer.

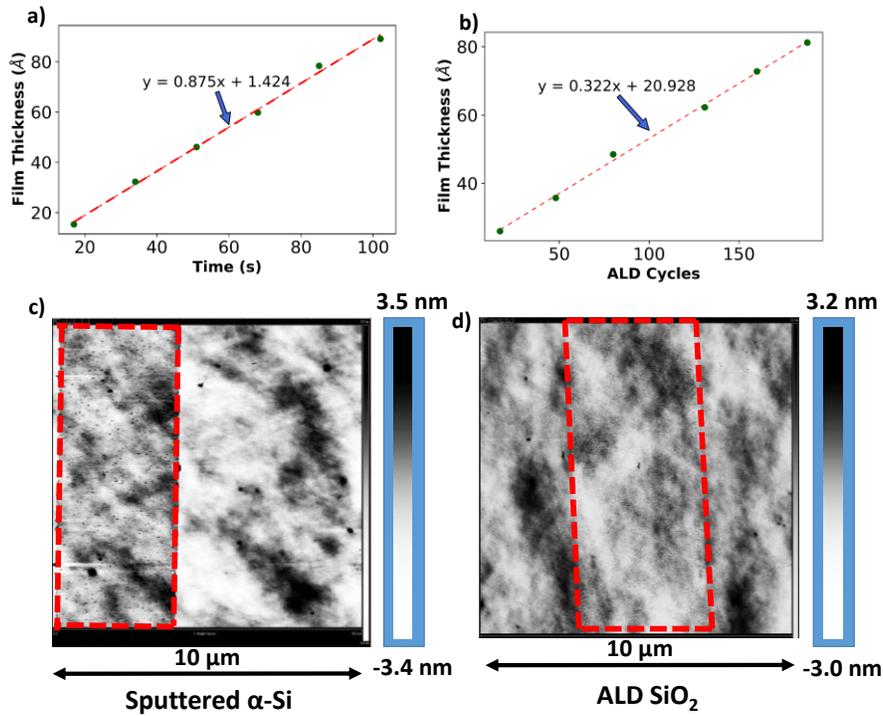

**Figure 2.3**: Thickness calibration curves for (a) sputtered α-Si and (b) ALD SiO$_2$ and (c) and (d) Modified AFM scans showing surface roughness of α-Si ALD SiO$_2$ thin films. The red dotted lines outline the features patterned using conventional lithographic techniques.

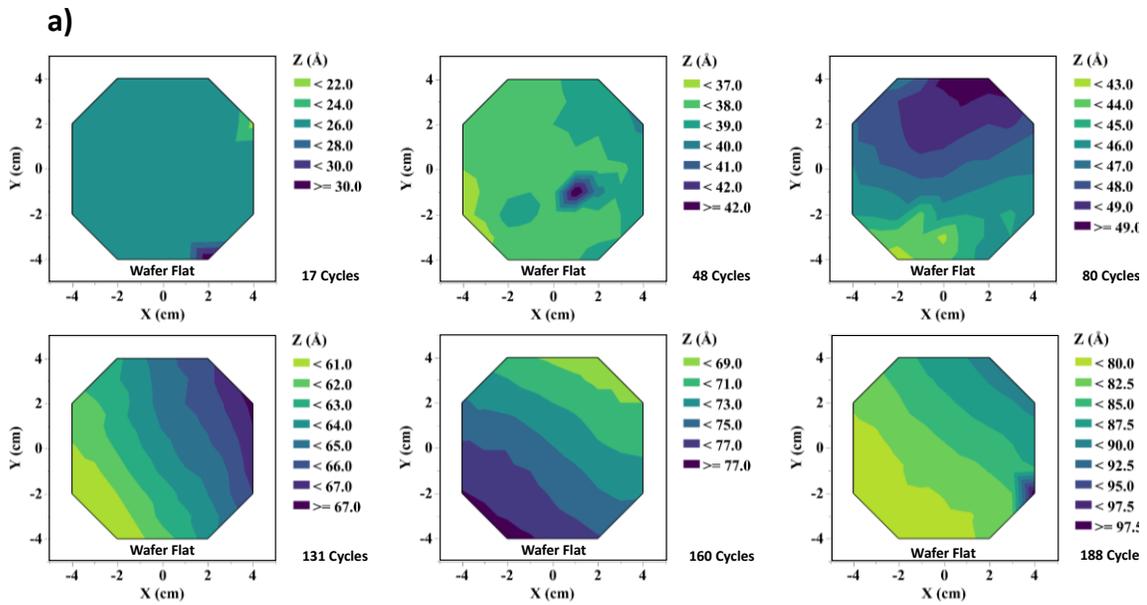

b)

| No. of Cycles (#) | Std. Dev of Sample 1 (Å) | Std. Dev of Sample 2 (Å) | Std. Dev of Sample 3 (Å) | Std. Dev of Sample 4 (Å) | Std. Dev of Sample 5 (Å) | Average Std. Dev (Å) |
|---|---|---|---|---|---|---|
| 17 | 2.51 | 1.07 | 3.59 | 2.95 | 0.98 | 2.22 |
| 48 | 0.86 | 9.49 | 0.91 | 8.59 | 9.56 | 5.88 |
| 80 | 3.9 | 1.74 | 5.95 | 4.02 | 9.32 | 5.23 |
| 131 | 5.62 | 4.63 | 1.93 | 2.77 | 7.35 | 4.46 |
| 160 | 3.23 | 6.52 | 2.70 | 3.15 | 3.23 | 3.77 |
| 188 | 8.5 | 5.61 | 3.70 | 8.18 | 4.20 | 6.04 |

**Figure 2.4**: Film thickness contour mapping, as measured by optical methods, on 4-inch wafers for different thicknesses of SiO2 and nonuniformity measurements of different SiO2 films done on five samples for each specific thickness.

Gogoi et al., [1,2], also developed batch-fabricated multisensor platforms on a single chip which are probable candidates to be used for IoT based applications. Molecular devices based on this construction method have been previously used as tunneling chemiresistors for bio sensing [3–12] and are potential candidates for low-power consumption gas sensing devices [9,13–20]. Although in the last couple of years, we have witnessed novel methods of stiction removal, compatible to conventional MEMS processes [21,22].

## Conclusions

The aim of this research was the development of a new class of chemiresistors which consume extremely low power during device operation, are highly sensitive and display low cross-sensitivity towards commonly found VOCs and gases. In accordance with these aims, we fabricated two new types of gas-sensors. The first was a nanogap device, which was designed such that the functionalized device was exposed to the analyte, the sensor would trap the intended analyte molecules within the nanogap and the junction resistance would reduce by several orders of magnitude while only consuming 15 pW of DC power during "stand-by" mode. The second type of sensor was a quantum tunneling hygrometer which displayed several orders of resistance change when exposed to water vapor molecules. The humidity sensor consumed 0.4 pW of power during "stand-by" mode. Both the devices were batch-fabricated and are compatible with existing CMOS technology. They are therefore suitable for IoT based applications and low power sensing. The significant contributions of this work are listed below.

- Nanogap electrodes with spacer thicknesses as low as ~4.0 nm were fabricated and we performed extensive electrical characterization of the fabricated devices across a 4-inch wafer. A novel use of α-Si was also demonstrated as a high resistance adhesive layer for gold.

- Using standard optical techniques, extensive thickness uniformity characterization of the deposited ultra-thin films and process repeatability measurements were performed.
- Tunneling current measurements were performed on devices across the wafer and nonuniformities in I-V characteristics were investigated. Transition Voltage Spectroscopy method was used to determine the barrier potential of the spacer film.
- Breakdown measurements were also performed to determine the maximum operating voltage of these devices. Temperature response of the device was also monitored by measuring the I-V characteristics of the device at different substrate temperatures.
- The fabricated devices were functionalized with a fully conjugated ter-phenyl linker molecule, (4-((4-((4 mercaptophenyl) ethynyl) phenyl) ethynyl) benzoic acid) for electrostatic capture of our target gas -cadaverine.
- We demonstrated ultra-low power resistance switching in batch-fabricated nanogap junctions upon detection of target analyte - cadaverine. The stand-by power consumption was measured to be less than 15.0 pW and the $R_{OFF}/R_{ON}$ ratio was more than eight orders of magnitude when exposed to ~80 ppm of cadaverine.
- A phenomenological electrical model of the device is also presented in good agreement with experimental observations.
- Cross-sensitivity of the gas sensor was tested by exposing the device to a variety of the commonly found VOCs and other atmospheric gases. The experiments revealed a highly selective sensor action against most of these analytes.
- These batch-fabricated sensors consume ultra-low power and demonstrate high

- selectivity; therefore, they can be suitable candidates for sensor applications in power-critical IoT applications and low power sensing.
- The design, fabrication, electrical characterization and working of a temperature compensated tunneling humidity sensor was presented. Sensor response showed a completely reversible reduction of junction resistance of ~four orders of magnitude with a standby DC power consumption of ~0.4 pW.
- Passive temperature compensation was also achieved using a nondifferential swelling design for the polymer patches. Temperature response showed a reduction of ~2.5 times when the device was exposed to a temperature sweep of 25°C – 60°C, which is 0.0025% of the maximum sensor output when exposed to rising levels of humidity.
- A mathematical model and equivalent electrical circuit was also presented to accurately describe the current conduction mechanism responsible for sensor action. Finally, sensor response dynamics was also investigated and established sorption analytical models were used to describe the time dependent sensor action.

The fabricated devices represent a new type of ultra-low power chemiresistive sensors. Since they are first-generation devices, they demand improvements before they can be implemented commercially. Some of these are as follows:

- In order to improve the uniformity of the deposited thin films, advanced deposition techniques such as Pulsed Laser Deposition can be used during the fabrication process. This would also ensure a more selective sensing action based on the length of the analyte molecule.
- Since the molecules are captured in the nanogap present along the perimeter of the upper electrode in the nanogap device, an Auxetic-Fractal electrode design would ensure a higher number of capture sites whilst keeping the overlap area and by extension, foot-print of the device the same.

- Although we used a phenomenological electrical model to accurately describe sensor action which was in good agreement with the experimental data, the model was unsuccessful in providing quantitative measurements of energy levels of the individual molecules forming a bridge across the junction. In order to tackle this issue one must look towards using computational chemistry methods such as NEGF and DFT formulism to better quantify important junction characteristics of the molecular channel.
- Sensor recovery was a major issue that we faced during our experiments. This requires further characterization and assessment of sensor action and target molecules to enable enhanced recovery of the nanogap sensors.
- Although we did provide passive temperature compensation for our humidity sensor, experiments still showed a nonzero temperature response. Additionally, the device does not compensate for temperature-dependent polymer absorption behavior. Both these aspects can be addressed by improved designs to the humidity sensor.